\documentclass{article}
\usepackage{emulateapj}

\usepackage{graphicx}

\begin{document}

\newcommand{\ba}{\mbox{\boldmath{$\alpha$}}}
\newcommand{\bt}{\mbox{\boldmath{$\theta$}}}
\newcommand{\bb}{\mbox{\boldmath{$\beta$}}}

\title{Gravitational Shear, Flexion and Strong Lensing in Abell 1689}

\author{Adrienne Leonard$^{1,2}$, David M. Goldberg$^1$, Jason L.
  Haaga$^1$, Richard Massey$^3$} \affil{$^1$ Department of Physics,
  Drexel University, $^2$ Institute of Astronomy, University of
  Cambridge, $^3$ California Institute of Technology}

\begin{abstract}

We present a gravitational lensing analysis of the galaxy cluster
Abell 1689, incorporating measurements of the weak shear, flexion, and
strong lensing induced in background galaxies. This is the first time
that a shapelet technique has been used to reconstruct the
distribution of mass in this cluster, and the first time that a
flexion signal has been measured using cluster members as lenses.
From weak shear measurements alone, we generate a non-parametric mass
reconstruction, which shows significant substructure corresponding to
groups of galaxies within the cluster. Additionally, our galaxy-galaxy
flexion signal demonstrates that the cluster galaxies can be well-fit
by a singular isothermal sphere model with a characteristic velocity
dispersion of $\sigma_v = 295\pm 40\ km/s $. We identify a major,
distinct dark matter clump, offset by 40$h^{-1}$kpc from the central
cluster members, which was not apparent from shear measurements
alone. This secondary clump is present in a parametric mass
reconstruction using flexion data alone, and its existence is
suggested in a non-parametric reconstruction of the cluster using a
combination of strong and weak lensing. As found in previous studies,
the mass profile obtained by combining weak and strong lensing
data shows a much steeper profile than that obtained from only
weak lensing data.

\end{abstract}
\keywords{cosmology:observations - galaxies:clusters:general -
  galaxies:photometry - gravitational lensing}

\section{Introduction}
\label{sec:Introduction}

Gravitational lensing studies provide a powerful tool for mapping out
the surface mass density of galaxies and clusters of galaxies
(e.g. Bradac et al. 2005ab, 2006; Natarajan \& Springel 2004; Diego et
al. 2005a; Cacciato et al. 2006; Abdelsalam et al., 1998). In
particular, cluster lensing studies provide constraints on both
structure formation models and the mean mass density of the universe.
Additionally, accurate knowledge of the mass distribution in clusters
enables better determination of the relationship between dark matter,
gas and galaxies within the cluster, and allows us to probe how well
light traces mass.  For example, there has been tremendous excitement
recently about maps from gravitational lensing of the distribution of
baryonic mass and dark matter in large-scale structure (Massey et
al. 2007a) and particularly the cluster 1E0657-56 (the ``Bullet
Cluster''; Bradac et al. 2006), in which there is a marked offset
between the peaks of baryonic and non-baryonic mass.

Traditional approaches to weak gravitational lensing focus on
estimating the mass of the lens by measuring the induced ellipticity
(shear) in images of background galaxies (see, e.g. Kaiser, Squires \&
Broadhurst, 1995 [hereafter KSB], for a review of the canonical
approach inverting ellipticities; Bartelmann \& Schneider, 2001,
provide an excellent review of the subject). Underlying these
approaches is the assumption that any variations in the lensing field
over the scale of the galaxy image may be neglected. Flexion has
recently been introduced as a means to describe higher order lensing
effects which arise as a result of small-scale variations in the
lensing field (Goldberg and Bacon, 2005 [hereafter GB05]; Bacon,
Goldberg, Rowe and Taylor, 2006 [hereafter BGRT], Goldberg \& Leonard,
2007 [hereafter GL07]). The flexion technique enables us to probe
structure on smaller scales than does a shear analysis, and is
particularly useful in galaxy-galaxy lensing studies.

In this paper we present several mass reconstructions of Abell 1689, a
rich cluster of galaxies at a redshift of $\sim 0.18$, using images
taken by the HST Advanced Camera for Surveys (ACS).  This cluster has
been well studied in the context of both weak and strong lensing, and
various parametric models for the mass distribution have been tested
(see, for example, Bardeau et al., 2005; Halkola, Seitz \& Pannella,
2006; Saha, Read \& Williams, 2006; Zekser et al., 2006; Umetsu,
Broadhurst, Takado \& Kong, 2005; Broadhurst et al., 2005a, Diego et
al., 2005b, Sharon et al, 2005; King et al., 2002; King, Clowe \&
Schneider, 2002; Dahle, Kaiser, Squires \& Broadhurst, 2001; Dye et
al., 2001; Taylor \& Dye, 1998; Tyson \& Fischer 1995).  Recently,
Halkola, Seitz \& Pannella (2007) have investigated the size of
individual galaxy haloes within the cluster using strong lensing,
which is typically difficult to do in clusters using traditional weak
lensing techniques.

Likewise, a number of researchers, including the original observing
team (Broadhurst et al. 2005b; see also Diego et al. 2005b) have
produced parametric and non-parametric mass estimates using strong
lensing from the same ACS images in our sample. Most recently,
Limousin et al. (2006 [hereafter L06]) have used these images, as well
as images from the CFH12k camera, to model the cluster using both
strong and weak lensing measurements, and find good agreement in the
mass estimates derived from these two sets of measurements.

We present here a non-parametric mass reconstruction of this cluster
from shear measurements, as well as a galaxy-galaxy lensing study from
measurements of flexion. It is typically very difficult to extract a
weak shear galaxy-galaxy signal in clusters of galaxies, as the smooth
component of the cluster field dominates. However, flexion is better
suited to detect perturbations to this lensing field on smaller
scales, and thus is particularly adept at picking out lensing signals
from individual galaxies in the cluster.

We show that our flexion data can be used to construct a parametric
mass reconstruction that successfully identifies substructure within
the cluster. Additionally, we combine our non-parametric shear mass
reconstruction with a non-parametric strong lensing
reconstruction. The combined convergence map shows significant
substructure, and suggests the presence of the secondary dark matter
clump described by L06. Additionally, this convergence map shows a
steeper radial profile in the center of the cluster than the one
generated by shear measurements alone.

This paper is structured as follows: we begin in \S~\ref{sec:Theory}
with a brief review of flexion formalism and a discussion of our
measurement techniques, including an overview of the formalism
underlying the shapelets and HOLICs methods.  We describe our data and
processing pipeline in detail in \S~\ref{sec:Pipeline}, and present
the results of our study in \S~\ref{sec:Results}.

\section{Measuring Shear and Flexion}
\label{sec:Theory}

\subsection{Flexion Formalism}
\label{subsec:Flexion}

It is helpful to begin by specifying our notation.  Consider an image
(at $z=\infty$) which, in the absence of lensing would be observed at
angular position $\bb$.  This may be related to the observed
coordinate, $\bt$ via the 2nd order transformation:
\begin{equation}
  \label{eq:betatheta}
   \beta_i\simeq A_{ij}\theta_j +\frac{1}{2}D_{ijk}\theta_j\theta_k,
\end{equation}
where
\begin{eqnarray}
  {\bf A(\bt)} &\equiv& \frac{\partial {\bb}}{\partial {\bt}}\nonumber \\
& =& \left(
    \begin{array}{clrr}
      1-\kappa-\gamma_1 & \ \ \ -\gamma_2\\ -\gamma_2 & 1-\kappa+\gamma_1
    \end{array}
  \right)
\label{eq:A}
\end{eqnarray}
and
\begin{equation}
  D_{ijk} = \partial_k A_{ij}.
\label{eq:D}
\end{equation}
The ${\bf D}$ operators can be expressed as follows:

\begin{eqnarray}
  D_{ij1}= \left( \begin{array}{clrr}
  -2\partial_1\gamma_1-\partial_2\gamma_2 & -\partial_1\gamma_2\\
  -\partial_1\gamma_2 & -\partial_2\gamma_2
    \end{array}\right)\nonumber \\ 
  D_{ij2}=\left( \begin{array}{clrr} -\partial_1\gamma_2 & \ \ \
  -\partial_2\gamma_2\\ -\partial_2\gamma_2 &
  2\partial_2\gamma_1-\partial_1\gamma_2\end{array} \right).
\end{eqnarray}

These operators give rise to an asymmetrical distortion in the image,
namely a skewness and a bending or arciness, as well as a shift in the
centroid of the light distribution. BGRT define two ``flexions'' as
follows:

\begin{equation}
  {\cal F} = \partial^\ast\gamma =
  (\partial_1\gamma_1+\partial_2\gamma_2) +
  i(\partial_1\gamma_2-\partial_2\gamma_1)
  \label{eq:F}
\end{equation}
and
\begin{equation}
  {\cal G}=\partial \gamma = (\partial_1\gamma_1-\partial_2\gamma_2) +
  i(\partial_1\gamma_2+\partial_2\gamma_1),
\label{eq:G}
\end{equation}

where $\partial = \partial_1+i\partial_2$ is the complex gradient
operator defined in BGRT. First flexion, $\cal{F}$, has an $ m =1 $
rotational symmetry (i.e. it is a vector) and gives rise to a skewness
in the light distribution of a galaxy image, as well as a shift in its
centroid. It is also simply related to the convergence, $\kappa$, by
${\cal F} = \partial \kappa$.  Second flexion, $\cal{G}$, has an $m=3$
rotational symmetry, and gives rise to bending in the lensed image.

\subsection{Measurement Techniques}
\label{subsec:Techniques}

Two distinct methods have been proposed for measuring flexion in real
images. The method described in GB05 and BGRT involves the
decomposition of galaxy images into shapelets (Refregier 2003),
followed by an ``active'' perturbation of coefficients (adopting the
terminology of the Shear TEsting Programme; Heymans et al., 2006;
Massey et al., 2006). A ``passive'' method, involving high-order
moments of a galaxy's shape, referred to as Higher Order Lensing
Image's Characteristics (HOLICs), was first related to flexion by
Okura et al. (2007). These techniques have been refined, compared, and
discussed in detail in GL07.  We also note that others (Irwin \&
Shmakova 2005; 2006) have proposed a statistic similar to flexion,
using a $\chi^2$ minimization technique to fit a polynomial radial
profile with first- and second-order perturbations.

\subsubsection{Shapelets}
\label{subsubsec:Shapelets}

Several researchers (Refregier, 2003; Bernstein \& Jarvis, 2002) have
proposed techniques for decomposing images into ``shapelets'', simple
basis functions composed of the two-dimensional, Gaussian-weighted
Hermite polynomials $B_{nm}(\bt)$.  Any isolated image $f(\bt)$ can be
expressed as a sum
\begin{equation}
f(\bt)=\sum_{mn}f_{nm}B_{nm}(\bt)\ ,
\end{equation}
where the weights $f_{nm}$ are known as ``shapelet coefficients''.  As
shown in Refregier (2003), Refregier \& Bacon (2003), Massey \&
Refregier (2005) and Massey et al. (2007b), shapelets are an
especially useful basis for our purposes because image transformation
operations induced by gravitational lensing typically produce very
compact transfers of power between shapelet coefficients.

In the shapelets framework, the lensing operators can be expressed
elegantly in terms of the quantum mechanical raising and lowering
operators, $a$ and $a^\dagger$, which are combinations of the
$\theta_i$ and $\partial_i$ operators.

We can express the lens equation as:
\begin{equation}
  f =
  (1+\kappa\hat{K}+\gamma_j\hat{S}_j+\gamma_{i,j}\hat{S}^{(2)}_{ij})f^\prime,
\label{eq:lens}
\end{equation}
where $f^\prime$ represents the unlensed image, $f$ refers to the
lensed image, and the various operators $\hat{K}$, $\hat{S}_j$ and
$\hat{S}^{(2)}_{ij}$ are expressed in terms of $a$ and $a^\dagger$
(the reader is referred to GB05 for the details).

An important feature of the operators in Equation \ref{eq:lens} is
that the $\hat{K}$ and $\hat{S}_j$ operators transfer power in
shapelet space such that $|\Delta n| +|\Delta m| = 2$, whereas the
second-order operators $\hat{S}^{(2)}_{ij}$ yield $|\Delta n| +
|\Delta m| = $ 1 or 3. This provides a straightforward mechanism for
extracting the first and second order signals.

As with a shear analysis, our flexion analysis makes the assumption
that any intrinsic flexion signal will be randomly oriented, and thus
will average to zero. This means that we expect that on average the
($n+m=$) odd shapelet modes will have a zero signal.

In practice, the shapelet series will need to be truncated, in which
case there are two parameters that must be specified for the
decomposition. These parameters are $n_{max}$, the maximum order of
the shapelet series, and $\beta$, the characteristic scale of the
shapelets:

\begin{equation}
{\cal B}_{nm}(\bt)\propto
\exp \left(-\frac{\theta_1^2+\theta_2^2}{2\beta^2}\right).
\end{equation}
As discussed in Refregier (2003), the optimal choices for $\beta$ and
$n_{max}$ are:
\begin{eqnarray}
\label{eq:betanmax}
\beta \simeq \sqrt{\theta_{max}\theta_{min}}\nonumber\\ n_{max} \simeq
\frac{\theta_{max}}{\theta_{min}}-1,
\end{eqnarray}
where $\theta_{max}$ and $\theta_{min}$ are the maximum and minimum
scales on which one expects to resolve structure in the image,
respectively. For the data used here, we find $\theta_{min} = 1.0$
pixels and $\theta_{max} = 1.5\sqrt{a^2+b^2}$ give good shapelet
reconstructions of our galaxy images, where $a$ and $b$ are the
semi-major and minor axes of the object as measured during source
extraction, respectively.

Images also need to be corrected for PSF and pixellisation effects,
which both tend to dilute the lensing signal. One of the advantages of
the shapelets technique is that an explicit PSF deconvolution, and an
integration within pixels, can be carried out prior to any parameter
estimation. Refregier \& Bacon (2003) give an explicit form for the
deconvolution, and Massey \& Refregier (2005) derive the form of the
integrals. We perform these tasks using the publicly available IDL
software from the shapelets web
page\footnote{http://www.astro.caltech.edu/$^{\sim}$rjm/shapelets/}.

\subsubsection{HOLICs}

We have discussed above how shapelets may be used to measure both the
shear and the flexion of a lensed image.  Historically, however, shear
has primarily been measured using combinations of weighted moments of
the image.  Likewise, the HOLICs technique proposed by Okura et
al. (2007), and subsequently refined in GL07, provides a
straightforward and fast way of estimating flexion using weighted
moments of the light distribution of galaxies.  Okura et al. define
the following complex terms:

\begin{equation}
  \zeta \equiv \frac{(Q_{111}+Q_{122})+i(Q_{112}+Q_{222})}{\xi}
\end{equation}
and
\begin{equation}
  \delta \equiv \frac{(Q_{111}-Q_{122})+i(3Q_{112}-Q_{222})}{\xi},
\end{equation}
where
\begin{equation}
  \xi \equiv Q_{1111}+2Q_{1122}+Q_{2222}.
\end{equation}
As pointed out in GL07, the relationship between the above quantities
and flexion estimators is best expressed as a matrix equation:
\begin{equation}
  \label{eq:matrix}
  {\cal M} \left( 
    \begin{array}{clrr}
      {\cal F}_1\\{\cal F}_2\\{\cal G}_1\\{\cal G}_2
    \end{array}
  \right) = \left(
    \begin{array}{clrr}
      \zeta_1\\\zeta_2\\\delta_1\\\delta_2 
    \end{array} 
  \right),
\end{equation} 
where {$\cal M$} is a 4 x 4 matrix whose elements are proportional to
linear combinations of $Q_{ijkl}$ and $Q_{ij}Q_{kl}$. The reader is
referred to GL07 for the explicit forms of these elements.

Equation \ref{eq:matrix} describes how flexion could be estimated from
unweighted moments in the absence of any measurement noise.  However,
when dealing with real images, and particularly those dominated by sky
noise, measurement of these moments is inherently noisy. It is thus
necessary to include a weighting function in our measurements of the
moments, which gives higher weighting to the central region of the
postage stamp of a galaxy (where the actual image lies) than to the
extreme regions (which we expect to be dominated by noise).

We use a Gaussian filter:

\begin{equation}
  W(\bt) = \frac{1}{2\pi\sigma_{W}} \exp \left(-\frac
  {\theta_1^2+\theta_2^2}{2\sigma_W^2}\right),
\end{equation}
and define the weighted moments as:

\begin{equation}
  \hat{Q}_{ij} = \frac{1}{\hat{F}}\int (\theta_i-\overline{\theta_i})
  (\theta_j-\overline{\theta_j})f(\bt)W(\bt)d^2\bt.
\end{equation}

Using the weighted moments necessitates two corrections to Equation
\ref{eq:matrix}. The first of these corrections results from the fact
that there is a difference in the centroid shift between the weighted
and unweighted calculations. The second correction has to do with the
fact that the total flux is not conserved by lensing. Ordinarily, this
is related by the Jacobian of the coordinate transformation, however
when using a window function, the transformation must be considered
explicitly.

Incorporating these corrections, we can write

\begin{equation}
  \hat{\cal M} = {\cal M}(\hat{Q}_{ij},...)+\Delta{\cal M},
\end{equation}

where $\Delta{\cal M}$ is a 4 x 4 matrix representing the correction
terms. These terms are proportional to sums of $Q_{ijklmn}$ and
$Q_{ij}Q_{klmn}$. The reader is referred to GL07 for details regarding
the computation of these terms, which are expressed in full in the
Appendix to that paper, and available as an IDL function from the
flexion
website\footnote{http://www.physics.drexel.edu/$^\sim$goldberg/flexion/}.

\section{Data and Processing Pipeline}
\label{sec:Pipeline}
\subsection{Data}
\label{subsec:Data}

Our data consists of 20 HST ACS images of Abell 1689, taken using the
Wide Field Camera (WFC) during HST cycle 11 by H. Ford. These images
are all 2300-2400 second exposures covering a square field of view of
$3.4'$ on each side (corresponding to $\sim 460h^{-1}$kpc at the
distance of the cluster), and are described in detail in Broadhurst et
al. (2005b).  Of these images, 4 were taken using the F475W filter
(G-band), 4 using the F625W filter (R-band), 5 using the F775W filter
(I-band) and 7 using the F850LP filter (Z-band).

\begin{figure}[h]
\centerline{\includegraphics[angle=0,height=2.5in]{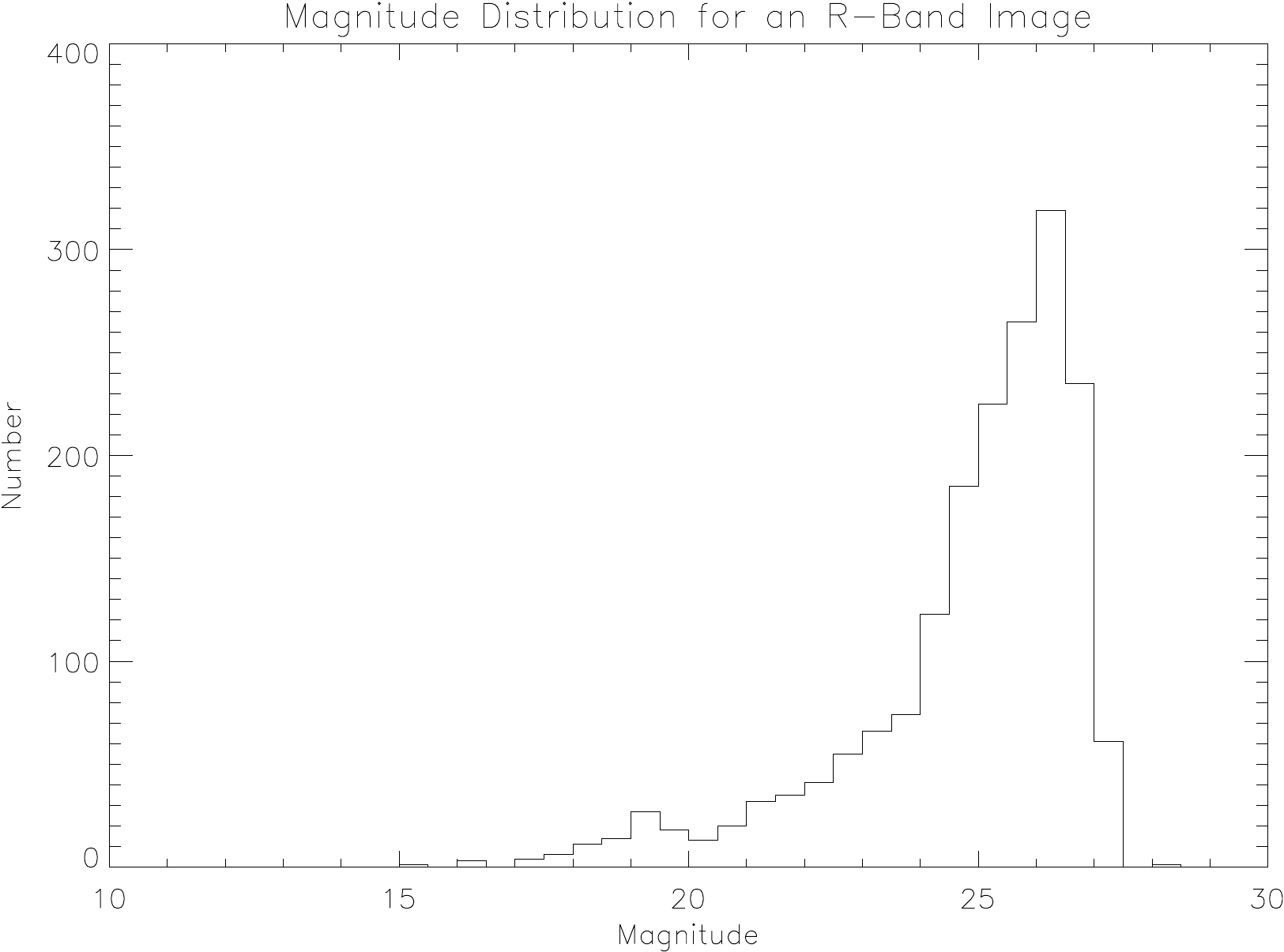}}
\caption{The figure shows the distribution in magnitudes of galaxies
  in an R-band image of Abell 1689. There are two clearly identifiable
  peaks in this distribution. While the majority of the objects in our
  sample are faint, with a magnitude peak around 26, there is also a
  low magnitude peak at around 19.  This corresponds primarily to
  cluster members, which are concentrated towards the center of the
  images, and bright stars in the field}
\label{fg:maghist}
\end{figure}

Due to the sensitivity of the ACS camera, these exposures resolve
objects down to a magnitude of $\sim 27$ in the R-band, which makes
them ideal for looking at very faint background sources.
Additionally, this sensitivity gives rise to a very high background
source count: our final catalog had $\sim 200$
sources/arcmin$^2$. However, in images of Abell 1689, the central
cluster galaxies are a dominant feature, with R-band magnitudes as low
as 15-16. Figure~\ref{fg:maghist} shows the magnitude distribution for
sources in an R-band image, and two distinct peaks can be identified.
The lower magnitude peak corresponds primarily to central cluster
members and bright stars, while the background galaxies are seen in
the larger, high-magnitude peak.
\clearpage

The large spread in magnitudes seen in Figure~\ref{fg:maghist}
presents several difficulties in the source detection and extraction
process. Typical source extraction software packages, such as
SExtractor (Bertin \& Arnouts, 1996), require various detection and
object size thresholds to be set prior source extraction. In images
where there is such a large range of brightnesses, setting lenient
detection criteria will result in detection of fainter background
sources, but will also lead to excessive blending of images in the
neighbourhood of the brightest sources. On the other hand, a strict
set of criteria will reduce the amount of blending seen, but will also
favor detection of foreground objects over the background objects that
we are interested in. In \S~\ref{subsubsec:SE} we discuss the
implementation of a two-pass source detection strategy carried out
using SExtractor.  This technique is modelled on the ``hot'' and
``cold'' source detection strategy described by Rix et al. (2004), who
first noted that a single set of SExtractor parameters is generally
insufficient for the detection of all objects of interest in ACS
images.

\subsection{Catalog Generation}
\subsubsection{Co-addition of Images}

\label{subsec:swarp}

For the initial source detection, we co-added our images using the
SWarp software package\footnote{http://terapix.iap.fr/soft/swarp}.
This software produces background-subtracted, median-stacked images,
which allows us to clean our images of spurious hotspots and bad
pixels. Images are read in by the software, and the background is
estimated and subtracted out. The images are then resampled and
projected into subsections of the output frame using any of a number
of astrometric projections defined in the WCS standard. Of the
projections included in the software, we opted to use the distorted
tangential (gnomonic) projection. This projection introduces very
little distortion in images smaller than 10 degrees (see Goldberg \&
Gott, 2006), and is recommended in the SWarp software manual for small
fields. The images are then co-added by taking the median value at any
given location.

In addition to cleaning our images of bad pixels, stacking the images
increases the signal-to-noise within a given image, thus better
enabling us to detect faint background sources. A single stacked image
was generated in the G-, R- and I-bands. In the Z-band, we opted to
create two stacked images. Our initial dataset consisted of 7
exposures in this band. For the purposes of source detection, we found
that very little advantage was gained by stacking 7 images compared to
stacking 3 or 4 frames. Thus, we created two Z-band images,
comparisons between which provide an important test of the
effectiveness of our source detection and extraction strategy.

\subsubsection{SExtractor Two-Pass strategy}
\label{subsubsec:SE}

We use SExtractor (Bertin \& Arnouts, 1996) to identify and extract
sources in our images. This software allows the user to specify a
number of input and output parameters, most notably a detection
threshold and a minimum number of connected pixels required to be
considered an object. Additionally, the software is able to produce
background and variance maps for the input images, and to generate
background-subtracted frames.

As noted above, a single SExtractor run will generally result in an
incomplete catalog of background sources. Thus, source extraction is
carried out in two stages. The first stage is designed to detect only
foreground objects by employing the cross-identification utility
included in the software. This utility allows the user to supply a
catalog of object positions, and to instruct the software either to
include or exclude the sources in the catalog. We supplied the
locations of known foreground objects compiled from the list of
spectroscopically confirmed cluster members presented by Duc et
al. (2002), prominent stars identified by eye, and objects identified
as cluster members by the NASA/IPAC Extragalactic Database (NED). For
this pass, we set the detection threshold at 4$\sigma$, and the
minimum area of detected objects at 10 pixels.

During this first run, background and variance maps were generated,
and this information was used to mask out the foreground objects in
order to simulate an emptier field. A friends-of-friends algorithm was
used to identify connected pixels above our detection threshold at the
locations of the foreground objects. These pixels were then masked out
according to $f_{pixel} = f_{background}+{\cal{R}}\sigma_{pixel}$
where $\cal{R}$ is a random number drawn from a standard normal
distribution ($\mu = 0$, $\sigma=1$), $\sigma_{pixel}$ is the standard
deviation of the background measured at the pixel location, and
$f_{background}$ is the background level. Thus, the foreground objects
are replaced with a simulated noisy background which has the same
statistical properties as the background measured in that region of
the image.

A secondary SExtractor run was then performed on the masked image,
using less stringent detection criteria (detection threshold:
2$\sigma$, minimum area: 15 pixels), to pick out background galaxies
in the field. This strategy alleviated any blending problems we might
otherwise have had by removing foreground objects prior to lowering
the detection threshold. Additionally, it removed from our catalog
known foreground objects which, if included in the subsequent
analysis, could dilute the measured shear and flexion signals.

A final catalog of background objects was generated by comparing the
image detections across different bands. An object was included in the
catalog if it was detected in at least two different bands, one of
which was either the R- or I-band.

Whilst the stacked images allow us to better identify legitimate
background sources, they are generally not suitable for carrying out
measurements, particularly of shear. This is because the PSF varies
from image to image; each exposure is offset slightly from the others,
and this results in a very complicated PSF in the stacked image that
is impossible to model simply. For this reason, the shapelet
measurements were carried out on the unstacked images, allowing an
explicit PSF deconvolution to be carried out during the shapelet
analysis.

It was thus necessary to run SExtractor on the individual frames,
since the shapelet parameters used in the analysis depend on
SExtractor measurements of galaxy shapes in a given image
($\theta_{max}=1.5\sqrt{a^2+b^2}$). We used the two-pass strategy
described above, using detection thresholds of 1$\sigma$ less than
those used on the stacked images to reflect the lower signal-to-noise
in the individual frames (the minimum source areas remained the
same). Additionally, in the second pass, we required that SExtractor
detect only those objects that were detected in the stacked frames and
included in the master background object catalog. This requirement
avoided any spurious detections resulting from bad pixels within the
individual exposures.

\subsection{Shapelet Analysis and Correction of Image Distortions}
\label{subsec:decomp}

Source extraction was carried out on each individual frame, generating
a catalog for that frame that contained information about the size,
shape and magnitude of each detected object. The next task was to
generate a postage stamp of each individual galaxy, and perform a
shapelet decomposition in order to measure the shear and flexion.

To generate a postage stamp, a circular region around each galaxy
image was extracted. The size of this region was determined by the
SExtractor estimate of the size of the galaxy. A segmentation map was
then created by identifying all pixels belonging to the primary
source, as well as any other sources found within the extracted
region. At this stage, if the primary source was found to extend close
to, or beyond, the boundary of the postage stamp, the size of the
extracted region was increased appropriately. The segmentation map was
then checked for blending of the primary source. Where there was no
blending, any other sources in the region were masked out, using a
similar procedure to that used in masking out the foreground
objects. Postage stamps in which blending of the primary source was
found were automatically rejected.

The next step in the process is a shapelet decomposition and PSF
deconvolution. As discussed in GL07, the computing time required for a
shapelet decomposition scales roughly as $\theta_{max}^4$ (not
including a PSF deconvolution, which slows the process further). This
means that for very large objects, the decomposition time becomes
prohibitive. GL07 discuss a method for speeding up the decomposition
in larger objects by re-gridding them into larger pixels. This method
works adequately, however it does not take into consideration any PSF
effects. We discuss the effect of the PSF on shear and flexion
measurements below, and describe our strategy for minimizing the
shapelet analysis time without significant reduction in the accuracy
of our shape measurements.

\subsubsection{PSF Modelling Using TinyTim}

\label{subsubsec:tinytim}

We modelled the PSF using the TinyTim software package (Krist,
1993). This software was designed to simulate the PSF of the WFC using
a three-step process that allows it to account for variations in the
PSF due to chip location and filter wavelength. We used the default
WFC settings for the focus and PSF size in our TinyTim simulations.

The ACS wide field camera has a well-studied geometric distortion in
its images, which is a result of the off-axis location of the camera
on the Hubble Space Telescope. This distortion has been modelled
(Meurer et al., 2003), and is well fit by a $4^{th}$ order
polynomial. TinyTim attempts to take this distortion into account by
applying it to the PSF in the third stage of the PSF generation.

This is necessary when correcting for the PSF at the point of
parameter estimation. However, our shapelet technique involves an
explicit PSF deconvolution before any shape measurements are
made. Including this correction to the PSF prior to deconvolution does
not amount to correcting the image for the geometric distortion. Thus
we did not include this step in our PSF generation; rather, we used an
undistorted PSF model in the deconvolution. At the point of parameter
estimation, we applied a correction for the geometric distortion as
described in \S~\ref{subsec:geom}.

As the decomposition time in larger objects is of concern, one might
ask what effect this model PSF has on shear and flexion
measurements. The flexion measured in the PSF is small (typically of
order $10^{-4} - 10^{-5}$/pixel). This will induce a flexion in the
image which scales roughly as:
\begin{equation}
  {\cal F}_{induced} \sim {\cal
  F}_{PSF}\frac{a_{PSF}^4}{a_{source}^4+a_{PSF}^4},
\end{equation}

where $a_{PSF}$ and $a_{source}$ refer to the semi-major axis of the
PSF and the source, respectively. This scaling relation follows from
equations 43-45 in GL07, and is discussed in more detail in
Appendix~\ref{sec: derivation}. Clearly, this drops off very rapidly
with increasing source size, thus the PSF will be a sub-dominant
effect in measurements of flexion for images with a semi-major axis
comparable to that of the PSF, which, for a typical R-band simulation,
is found to be $a_{PSF} \simeq 2.0$ pixels. Thus, the flexion
measurements, in general, do not need to be corrected for PSF effects,
particularly if a minimum size criterion is introduced (as in
\S~\ref{subsec:gg}).

In order to assess how the PSF affects the measured ellipticities of
galaxies, we simulated gaussian ``galaxies'' of various sizes and
ellipticities, convolved them with a model R-band PSF in real space,
and computed the change in ellipticity of the source. The ellipticity
was computed using the unweighted moments of the light distribution,
and we defined the change in ellipticity of the source as:
\begin{equation}
  \label{eq:deltae}
  \Delta \mid\epsilon\mid \equiv
  \sqrt{(\epsilon_1^{(0)}-\epsilon_1^{(c)})^2 +
  (\epsilon_2^{(0)}-\epsilon_2^{(c)})^2}
\end{equation}
where the superscript $(0)$ and $(c)$ refer to the unconvolved and
convolved images, respectively.

\begin{figure}[h]
  \centerline{\includegraphics[angle=0,height=2.5in]{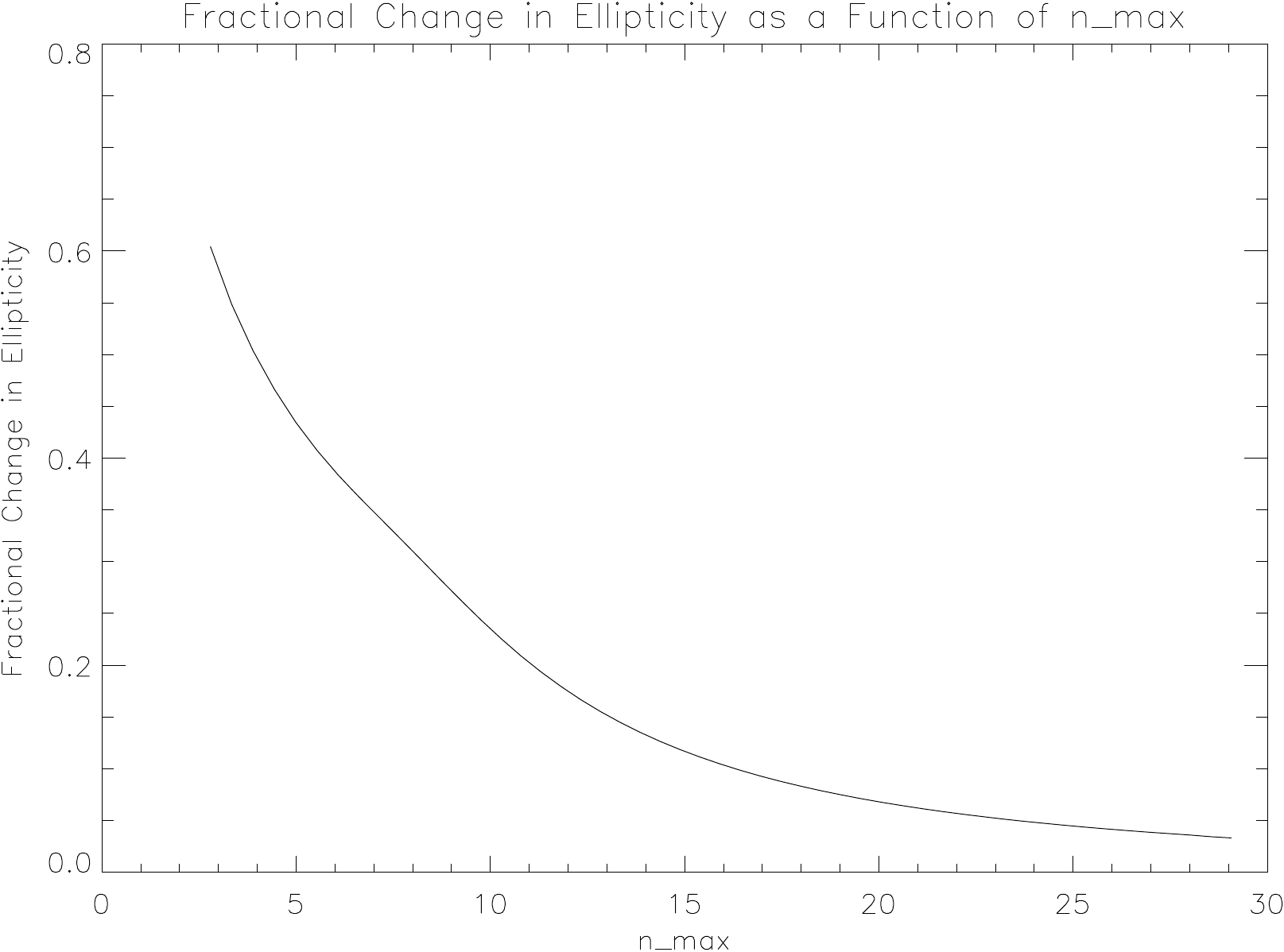}
    \includegraphics[angle=0,height=2.5in]{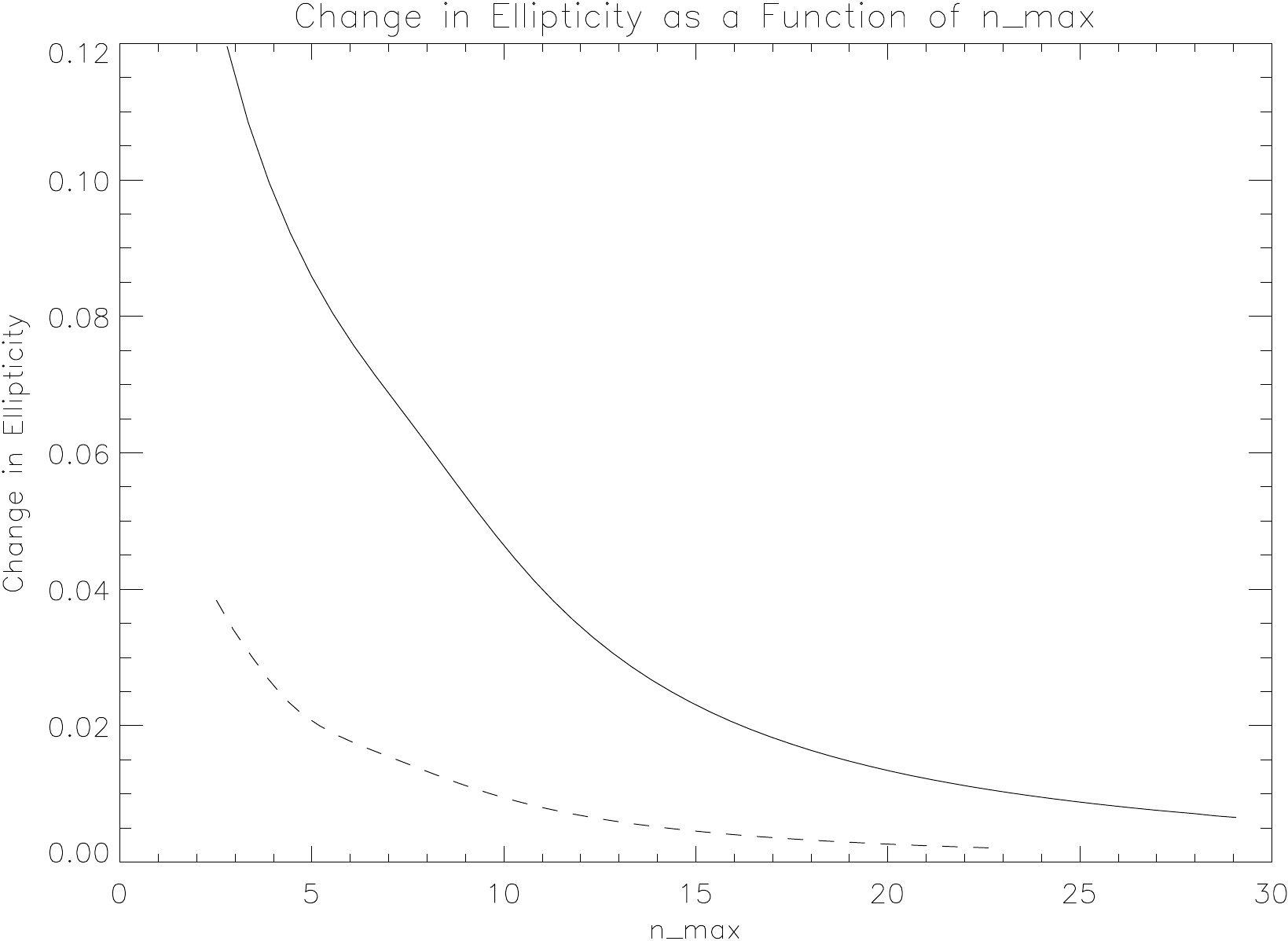}}
\caption{The left panel shows the fractional error in the ellipticity
    induced by the PSF plotted against
    $n_{max}=\frac{\theta_{max}}{\theta_{min}}-1$ for a series of test
    galaxies with an elliptical gaussian profile with $\mid \epsilon
    \mid = 0.2$. The right panel compares the absolute change in
    ellipticity for a circular source (dashed line) and an elliptical
    source with $\mid \epsilon \mid = 0.2$ (solid line).}
\label{fg:lgobj}
\end{figure}

The left panel of Figure~\ref{fg:lgobj} shows the fractional change in
ellipticity, $\frac{\Delta\mid\epsilon\mid}{\mid\epsilon\mid}$,
plotted as a function of $n_{max}$ (defined in
\S~\ref{subsubsec:Shapelets}) for an elliptical source with $\mid
\epsilon \mid = 0.2$. We also show in Figure~\ref{fg:lgobj} a
comparison of the absolute change in ellipticity for a circular and an
elliptical source.

From Figure~\ref{fg:lgobj}, we can see that the effect of the PSF on
the ellipticity of the source drops to below 10\% at $n_{max}=20$.
Thus we defined a ``large object cutoff'': any object with an $n_{max}
> 20$ was considered to be sufficiently large that PSF effects do not
dominate, and an explicit deconvolution was not carried out.
Additionally, as described in GL07, images with $n_{max} > 50$ were
re-gridded into larger pixels of size = $n_{max}/40$ in pixel units,
in order to further reduce the decomposition time for very large
objects. In order to exclude objects for which the shapelet fit was
not successful, we rejected any sources for which the decomposition
had a $\chi^2$ per degree of freedom that was greater than 2.0.

\subsection{ACS Geometric Distortion Correction}
\label{subsec:geom}

As described above, the ACS wide field camera induces a geometric
distortion in images that is well fit by a $4^{th}$ order polynomial
(Meurer et al., 2003):

\begin{eqnarray}
  \label{eq:geomfit}
  x^\prime = \sum_{i=0}^4\sum_{j=0}^i a_{ij}x^jy^{i-j}\nonumber\\
  y^\prime = \sum_{i=0}^4\sum_{j=0}^i b_{ij}x^jy^{i-j},
\end{eqnarray}
where the primed coordinates refer to the undistorted frame, and the
unprimed coordinates are the distorted pixel coordinates relative to
the central pixel in each chip. $a_{ij}$ and $b_{ij}$ are the best fit
coefficients, and are specific to each of the two chips on the ACS
camera. The fit parameters can be found in the TinyTim software package.

We can use these transformations to evaluate the {\bf A} and {\bf D}
operators defined in Equations \ref{eq:A} and \ref{eq:D}. This allows
us to evaluate the shear and flexion induced by the geometric
distortions as a function of location on the chip. These induced
distortions can then be subtracted from the measured shear and
flexion.

We found that the mean magnitude of the induced ellipticity across the
field of view was significant, with a mean of $\overline{\mid \gamma
\mid} = 0.0405$, and a standard deviation of $\sigma_{\mid\gamma\mid}
= 0.0083$. The induced flexion, however, was found to be negligible
over the entire field of view, with $\overline{\mid{\cal F}\mid} =
3.05 \times 10^{-4}$, $\overline{\mid{\cal G}\mid} = 1.00 \times
10^{-4}$, $\sigma_{\mid{\cal F}\mid} = 4.61 \times 10^{-5}$ and
$\sigma_{\mid{\cal G}\mid} = 2.34 \times 10^{-5}$.

In light of this, only the measured ellipticity was corrected for the
geometric distortions. Additionally, since the PSF was found to have
little effect on the flexion for images with $a \simeq 2.0$, we opted
to measure the flexion on the stacked images, for which the
signal-to-noise is higher, and incorporated a lower limit on the size
of images analyzed. We note that, as a result of the re-projection of
the images during the stacking process, the geometric distortions are
reduced in the stacked frames.

\subsection{Non-parametric Convergence Map Generation}

Each background galaxy had from one to twenty independent ellipticity
estimates, depending upon how many of fields it was detected in.  For
those detected at least twice, we estimated the mean and measurement
error of the galaxy ellipticities via the relation:
\begin{equation}
\sigma_{meas}=\frac{\langle (\varepsilon_i-\overline{\varepsilon})^2
 \rangle^{1/2}}{\sqrt{N-1}}.
\end{equation}
The mean and error were actually measured iteratively, and all
measurements outside of $2.5\sigma$ of the mean were discarded as
outliers.  An ellipticity error was then assigned to each galaxy as a
quadratic sum of the measurement uncertainty and $0.3$, the intrinsic
standard deviation in ellipticities.

We then binned the field into cells approximately 20 arcseconds on a
side with an estimated ellipticity given by the weighted mean of the
measurements in the individual frames, and with bin errors estimated
using standard error propagation. Various rejection criteria in the
pipeline (most notably blended sources and those with poor shapelet
reconstructions or fewer than two detections) resulted in a reduced
source count of 88/arcmin$^2$, which is much higher than that
typically found in weak lensing studies of this cluster (see, for
example, Broadhurst et al., 2005a).

We then applied the finite inversion technique advocated by Seitz \&
Schneider (1995).  They show that the density field may be solved
iteratively (up to the mass sheet degeneracy) as:
\begin{equation}
\kappa(\bt)-\kappa_0=\frac{1}{\pi}\int d^2\bt [1-\kappa(\bt')]{\cal
    R}[{\cal D}^\ast(\bt-\bt')\langle \epsilon\rangle(\bt')],
\end{equation}
where the convolution function ${\cal D}$ can be written as:
\begin{equation}
{\cal
  D}(\bt)=-\frac{\left[1-\left(1+\frac{\theta^2}{\theta_s^2}\right)
  \exp\left(\frac{\theta^2}{\theta_s^2}\right)
  \right]}{(\theta_1-i\theta_2)^2},
\end{equation}
and our smoothing scale, $\theta_s$, was set to half a binsize.  

The above method produces a reasonable mass reconstruction (see
\S~\ref{sec:Results}), however a better-constrained mass model can be
derived from combining both weak and strong lensing methods (see, for
example, Kneib et al., 2003; Smith et al, 2005; Bradac et al., 2005ab;
2006). In order to do this, we follow the method of Bradac et
al. (2005a). Our multiple image catalog is taken from L06, and we
assume that $D_{ls}/D_{s}=0.625$, i.e. the mean redshift of the
background sources is 0.9, as in L06.  Note that applying the Bradac
et al. technique with only weak lensing constraints produces a very
similar reconstruction to that found using the approach of Seitz \&
Schneider (1995).

\subsection{Parametric Convergence Map Generation}
\label{subsec:parametric}

Flexion typically dominates the lensing signal on a much smaller scale
than shear.  Thus, the binning scale described above is not
appropriate for flexion measurements as the signal would be entirely
dominated by shot-noise.  In this context, discrete estimates of the
flexion of individual galaxies (parametric reconstructions) are in
order.

As the flexion was measured on the stacked images, each source had up
to five flexion measurements. We included only those objects measured
in at least 3 of the 5 stacked frames, and combined the flexion
estimates using a similar iterative statistic to that described above
for shear.  Additionally, we only included objects that had a
semi-major axis larger than $0.12''$ (2.4 pixels), which were in the
brightest $90\%$ of the sample of background objects (which still
yielded approximately 75 galaxies/arcmin$^2$), and which had absolute
values of flexion that were smaller than $4\times$ the random scatter,
to exclude extreme outliers.

We found that the 2nd flexion produced significantly higher scatter
than the 1st flexion, and thus, 1st flexion was the only one used in
our analysis.

To generate a convergence map, we modeled each cluster galaxy under
the assumption that it is a singular isothermal sphere with the
flexion profile given in BGRT:
 
\begin{equation}
F_E(\theta)=\frac{-4\pi \left(\frac{\sigma_v}{c}\right)^2}{2\theta^2}.
\label{eq:flex_iso}
\end{equation}

The flexion data was used to fit $\sigma_v^2$ for each foreground
galaxy for which at least three background sources were identified
with an angular separation of $\le 10''$. Out of our initial sample of
62 foreground galaxies, only 36 met this criterion, and thus our final
mass reconstruction was based on the fits obtained for these 36
galaxies.

Since this is a noisy measure, some estimators predicted a negative
value of $\sigma_v^2$.  These were retained in our map in order to
keep the convergence estimate unbiased. The mean velocity disperson
computed this way was found to be $<\sigma_v>=321km/s$. After
estimating the velocity dispersions, a density field was layed down:
\begin{equation}
\kappa({\bt})=\sum_i 2\pi
\left(\frac{\sigma_{v,i}^2}{c^2}\right)\frac{1}{|\bt_i-\bt|}
\end{equation}
and then smoothed using a Gaussian kernel on a scale of $10''$.  

\subsection{Galaxy-Galaxy Flexion Measurements}
\label{subsec:gg}

We also used our flexion measurements in a galaxy-galaxy lensing
study. Galaxy-galaxy measurements of shear are extremely difficult in
cluster environments since a single source may be partially lensed by
many different foreground objects. Additionally, the shear signal
tends to be dominated by the smooth component of the cluster mass
distribution, rather than the substructure. The flexion signal,
however, drops off much more quickly with separation, and thus will
almost always be produced by a single source.

We carried out a pairwise comparison between each of the background
images and the 62 cluster members in our foreground object catalog.
For each, we considered the relative orientation angle and estimated
``B'' and ``E'' field flexion signals via the relationships:
\begin{eqnarray}
  {\cal F}_E \equiv \ \ {\cal F}_1\cos(\phi)+{\cal
    F}_2\sin(\phi)\nonumber\\
  {\cal F}_B \equiv -{\cal F}_1\sin(\phi)+{\cal F}_2\cos(\phi),
\end{eqnarray}
where $\phi$ is the position angle of the background source with
respect to the foreground galaxy. Lensing naturally gives rise to
${\cal F}_E<0$ and ${\cal F}_B = 0$, thus the scatter in the measured
B-mode signal gives an estimate of the noise in the measurements.

For each, we estimated an uncertainty in the flexion via the relation:
\begin{equation}
\sigma_F^2=\left(\frac{0.029}{a}\right)^2+\sigma_{meas}^2
\end{equation}
where the former term on the right represents the scatter in intrinsic
flexion (see GL07), and the latter represents the measurement error as
determined from the scatter between frames. Weighting each data point
by its signal-to-noise, we computed the average flexion signal as a
function of lens-source separation.

Finally, we performed least squares fitting to an assumed isothermal
sphere profile as described in Equation~\ref{eq:flex_iso}. Other fits
could be done, of course, but the isothermal model was both simple and
seemed to fit the data well. 

Note that, in this case, we are considering the average signal over
all foreground-background pairs, rather than fitting each individual
foreground galaxy.

\section{Results and Discussion}
\label{sec:Results}
\subsection{Weak Shear Mass Reconstruction (Non-Parametric)}

Most mass reconstructions of Abell 1689 have either been parametric
(e.g. L06; Halkola, Seitz \& Panella, 2006; Broadhurst et al., 2005b;
King, Clowe, \& Schneider, 2002), have measured circularly averaged
shears (e.g. L06; Bardeau et al., 2006; Broadhurst et al., 2005a;
Umetsu et al., 2005), or have been non-parametric reconstructions
using strong-lensing data (e.g. Diego et al., 2005b).  Thus, it is
interesting to compare weak non-parametric reconstructions to the
above methods to verify that they produce consistent results.  This
would also lend fuel to the recent efforts to unify weak and strong
lensing studies in order to further reduce uncertainty.

In order to generate radially averaged shear profiles of the cluster,
the authors mentioned above have made use of ground-based data, either
exclusively or as a supplement to the ACS images of the cluster. Part
of the reason for this is that the core of the cluster is quite
clumpy, and thus a centroid cannot be uniquely defined. That said, at
a distance of $100\arcsec$ from the brightest cluster member, we find
a tangential shear of about 0.2, in excellent agreement with the weak
lensing study of L06.

\begin{figure}[h]
\centerline{\includegraphics[angle=0,height=5in]{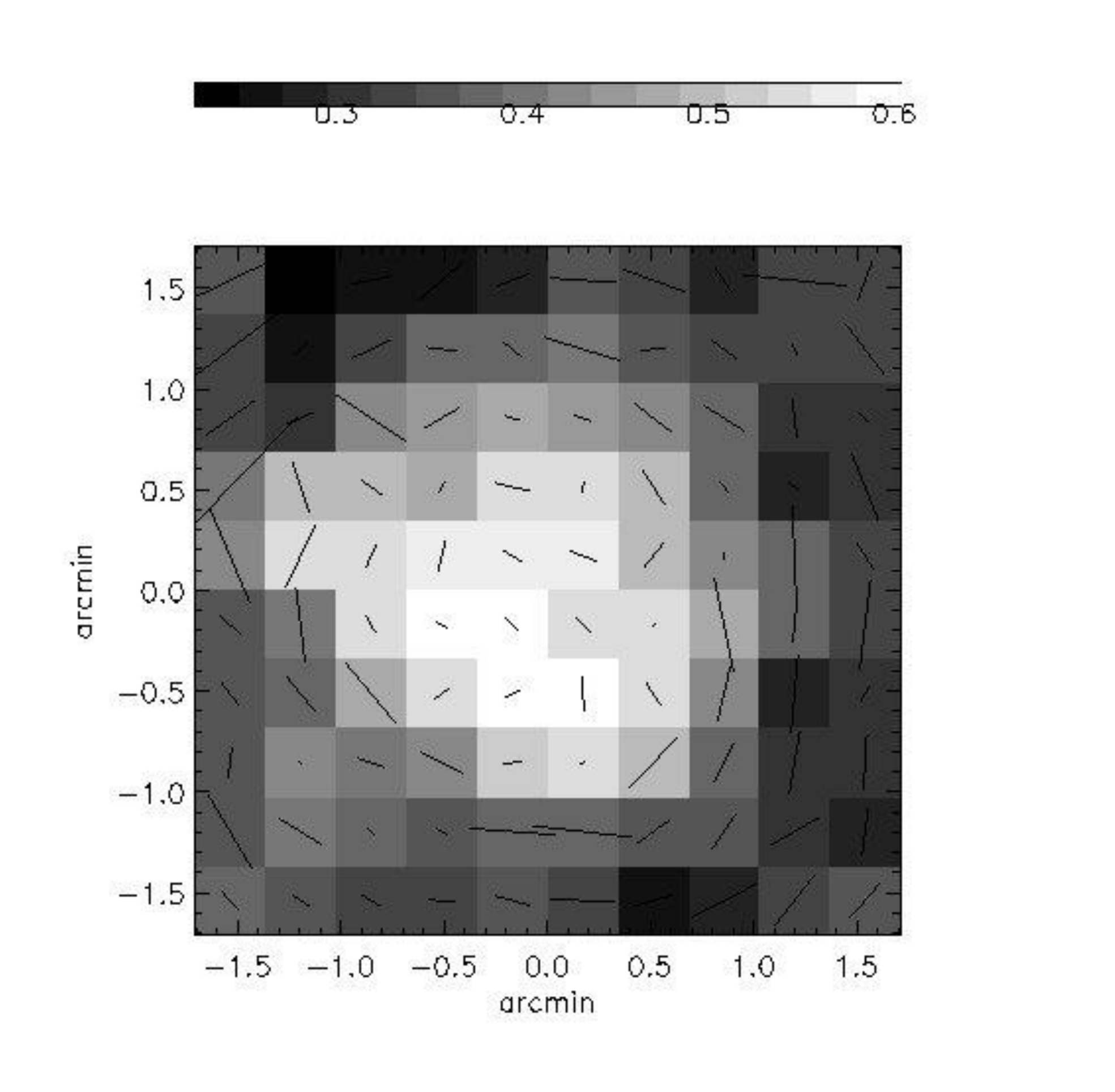}}
\caption{The weak-lensing shear field along with a non-parametric
  reconstrucion for ACS images of A1689. Note that the overall
  normalization of the reconstruction is subject to uncertainty due to
  the mass-sheet degeneracy in a finite field. North points to the
  lower right of the image, and the width of each bin corresponds to
  approximately $45h^{-1}$kpc at the distance of the cluster.}
\label{fg:shear_field}
\end{figure}

We have produced a simple shear field as described in the previous
section and used this field to reconstruct the density field via the
Seitz \& Schneider (1995) smoothed finite-inversion technique.  Our
shear field and reconstructed $\kappa$ field may be seen in
Figure~\ref{fg:shear_field}.

\begin{figure}[h]
\centerline{\includegraphics[angle=0,height=5in]{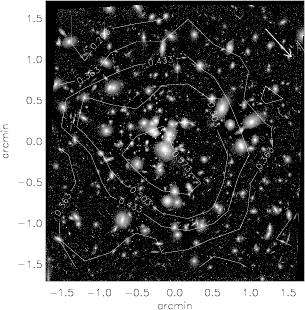}}
\caption{A contour plot of the convergence $\kappa$ determined
  using the shapelet measurements of the ellipticity of background
  sources in the field. The contours clearly identify small
  concentrations of foreground galaxies within the cluster.}
\label{fg:kappa}
\end{figure}

With so much attention given to the ability of strong lensing to pick
out substructure in clusters, it is gratifying to note that weak
lensing alone can identify large concentrations of galaxies.  A
contour plot of the projected surface mass density, $\kappa$, is shown
in Figure \ref{fg:kappa}, overlaid on an image of the cluster itself.

For this cluster, we find $\Sigma_{crit}\simeq 5.7 \times 10^{15} h
M_\odot Mpc^{-2}$ (taking $z = 0.18$), which allows us to estimate the
true surface mass density up to the mass-sheet degeneracy.  However,
since we have not independently measured the shear field far from the
center of the cluster, we rely on the estimates of others.  L06, for
example, found a mean $\overline{\kappa}$ of 0.48 within 1' of the
center of the cluster.  We have thus set our mass sheet at a slightly
lower threshold, so that the mean over the entire chip is 0.4.

We note that the density profile is quite shallow, however.  We find
that a power law profile is well fit if $\alpha=-0.19$.  Over a
similar range, L06 fit to a profile with $\alpha\simeq -0.6$.  L06,
however, base their fit on extrapolation from the slope of the shear
outside the ACS image.  Inspection of Fig. 11 in their work suggests
that the circularly averaged profile within the central arcminute may
be much shallower.

\subsection{Strong+Weak Lensing Mass Reconstruction (Non-Parametric)}

\begin{figure}[h]
\centerline{\includegraphics[angle=0,height=5in]{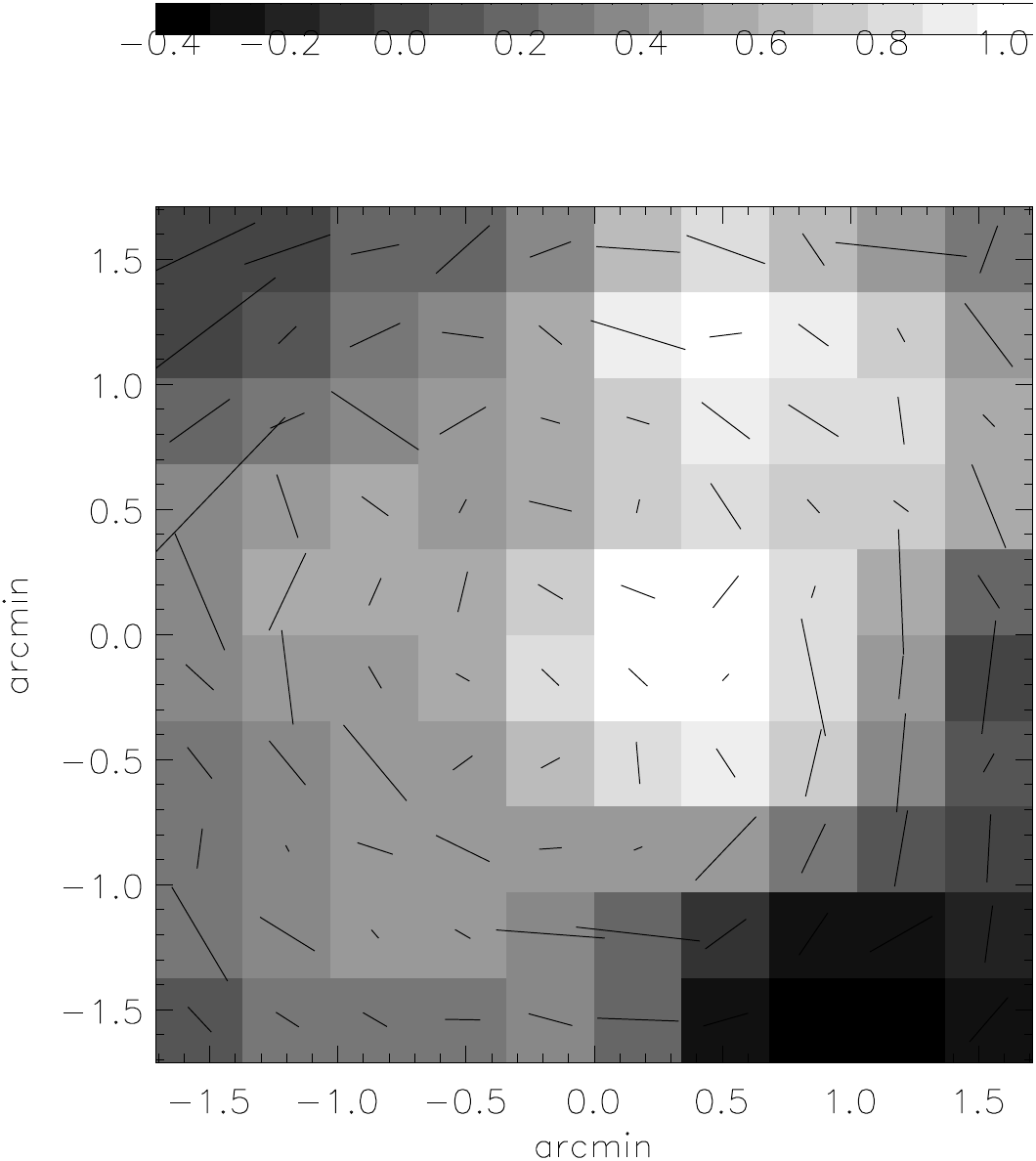}}
\caption{The shear map and the $\kappa$ field found using both
  shapelet measurements of the ellipticity and multiple image pairs
  identified by L06. In this plot, the width of each bin corresponds
  to $45h^{-1}$kpc in the cluster plane.}
\label{fg:S+W_shear}
\end{figure}

\begin{figure}[h]
\centerline{\includegraphics[angle=0,height=5in]{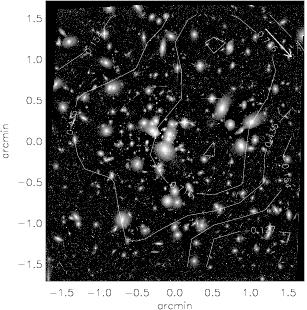}}
\caption{A contour plot of the convergence, $\kappa$, determined
  using the shapelet measurements of the ellipticity of background
  sources in the field combined with strong lensing data from
  L06.}
\label{fg:S+W}
\end{figure}

We have combined strong lensing information from L06 with our shear
measurements to generate a combined convergence map, following the
method of Bradac et al. (2005a). A convergence map and a contour plot
of the convergence found by this method can be seen in Figures
\ref{fg:S+W_shear} and \ref{fg:S+W}.

\begin{figure}[h]
\centerline{\includegraphics[angle=0,height=2.5in]{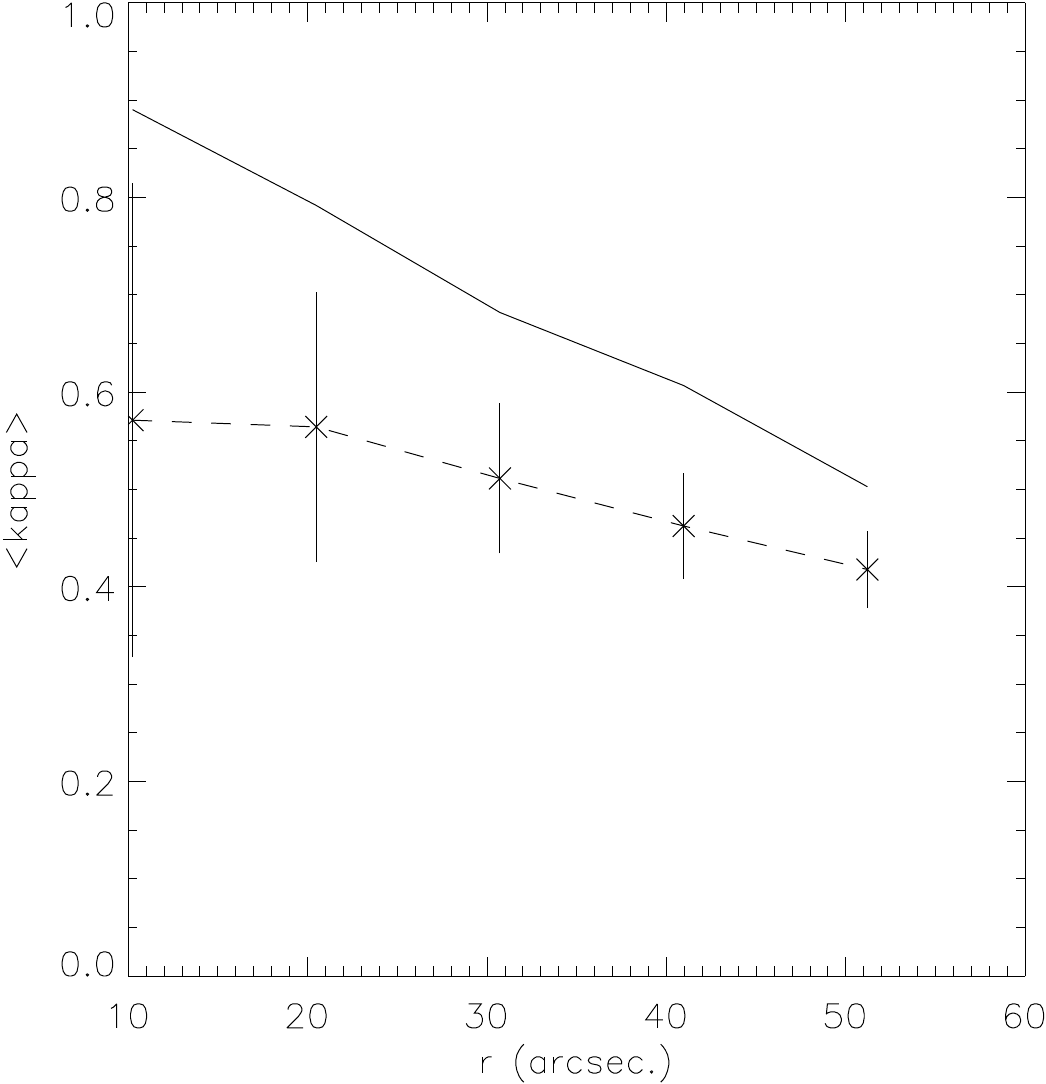}}
\caption{The circularly averaged convergence as a function of
  distance from the center of the cluster from weak measurements
  alone (dashed) and from the combined weak+strong calculation
  (solid line). Clearly, the weak+strong profile shows a much
  steeper slope than the weak profile. However, there is very good
  agreement between the two profiles at large radii. The error bars
  shown in these plots are taken from $\sqrt{N}$ statistics, thus
  are only an approximation.}
\label{fg:profiles}
\end{figure}

Figure \ref{fg:S+W} clearly shows an elongation in the direction of
the secondary dark matter clump described in L06, which is not seen in
Figure \ref{fg:kappa}. Figure \ref{fg:profiles} compares the
circularly averaged convergence as a function of distance from the
center of the cluster for the weak and strong+weak mass
reconstructions. The strong+weak profile has a much steeper slope than
the weak profile. Indeed, we find $\alpha=-0.29$ for the profile shown
in Figure~\ref{fg:profiles}.

This difference in slope could result from the fact that our masking
scheme results in an under density of background sources in the
central region of the cluster. This shortage of data points results in
a lower shear signal in the central region, thus lowering the computed
value of $\kappa$. The under density of sources in the central region
is found to be an important factor in the errors associated with our
parametric flexion reconstruction, and is discussed in more detail in
\S~\ref{subsec:flex_recomp}.

\subsection{Flexion Mass Reconstruction (Parametric)}
\label{subsec:flex_recomp}

A parametric mass reconstruction was generated using (first-) flexion
data alone. In dense systems like clusters, we have found that the
HOLICs technique is less susceptible than shapelets to contamination
by light from the extended wings of lens galaxies and other
neighboring sources. Flexion probes the higher order shape moments, so
even a small contamination near the edges of postage stamps can cause
a signicant spurious flexion signal.  We thus used our HOLICs
measurements exclusively to estimate the flexion signal.

\begin{figure}[h]
\centerline{\includegraphics[angle=0,height=5in]{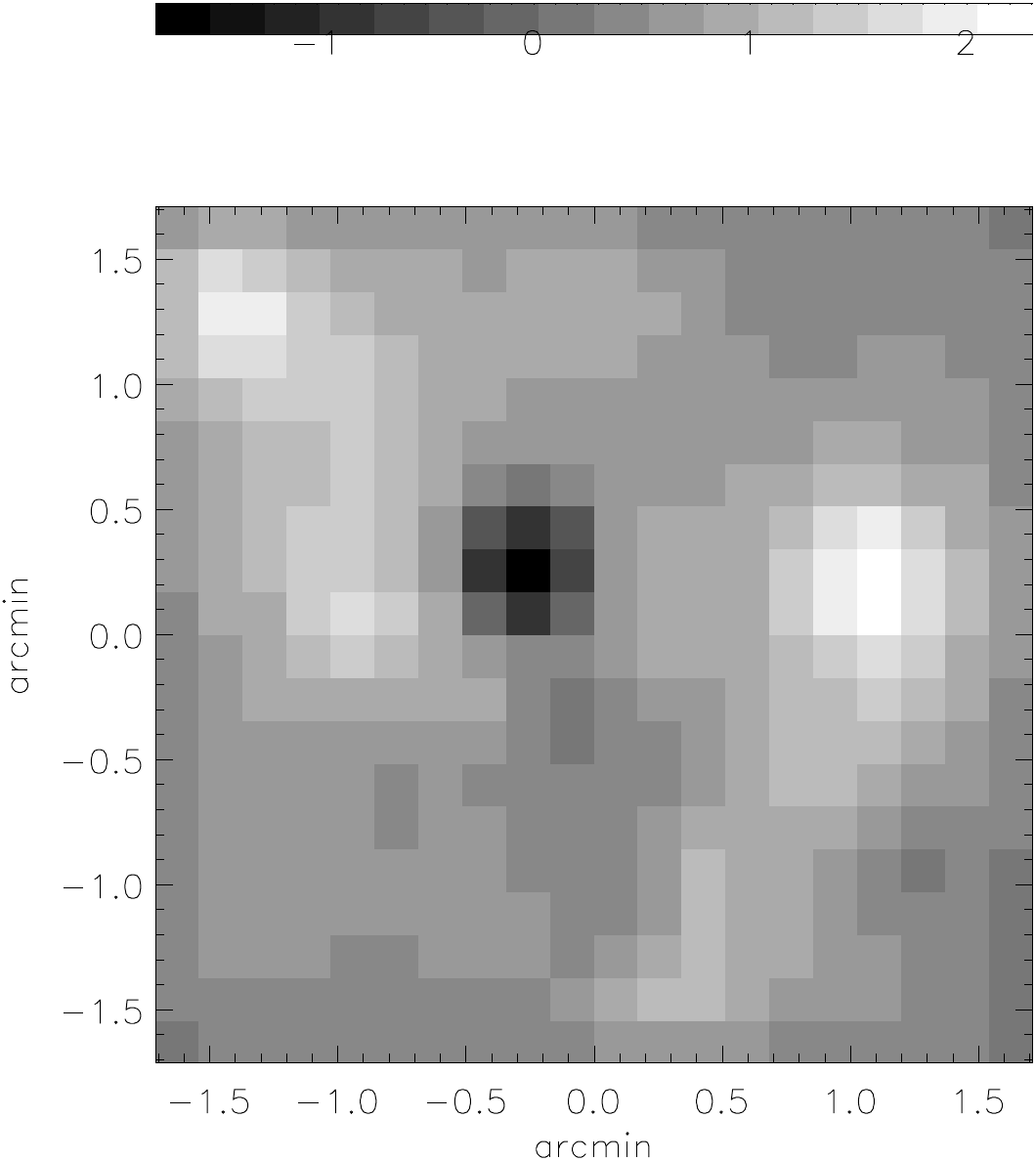}}
\caption{The binned convergence found using a parametric model for
  the cluster galaxies from flexion data. The width of each bin is
  approximately $22.5h^{-1}$kpc in the cluster plane.  Note that a
  flexion-only reconstruction is sensitive to localized substructure
  but not smooth gradients in the mass distribution.}
\label{fg:kappa_f}
\end{figure}

\begin{figure}[h]
\centerline{\includegraphics[angle=0,height=5in]{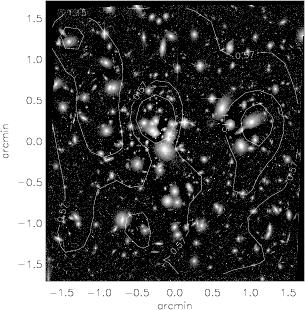}}
\caption{A contour plot of the convergence calculated using a
  parametric model for the cluster galaxies from flexion
  measurements. This reconstruction shows significant substructure
  corresponding to the locations of small clumps of cluster galaxies.}
\label{fg:contour_f}  
\end{figure}

The reconstructed convergence is plotted in Figure \ref{fg:kappa_f}
and a contour plot is shown in Figure \ref{fg:contour_f}. The
reconstruction shows significant substructure, which appears to be
well correlated with small clumps of galaxies outside the center of
the cluster. However, there is a rather worrying under density seen in
the center of the cluster. In order to asses the significance of this
under density, it is necessary to quantify the errors associated with
the convergence map.

\begin{figure}[h]
\centerline{\includegraphics[angle=0,height=5in]{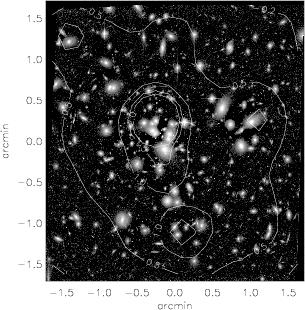}}
\caption{The figure shows a contour plot of the approximate errors in
  the parametric flexion mass reconstruction. It is evident from this
  plot that the reconstruction is entirely dominated by noise in the
  center of the image.}
\label{fg:error_f}
\end{figure}

Figure~\ref{fg:error_f} shows the approximate errors in the flexion
reconstruction. These errors were computed as follows: each flexion
data point was rotated by an angle drawn from a uniform random
distribution in the range $[0,2\pi]$. Parametric reconstructions were
generated using this randomized data, and this procedure was carried
out for 1000 randomizations. The errors presented in the figure are
the rms values found for each bin.

\begin{figure}[h]
\centerline{\includegraphics[angle=0,height=2.5in]{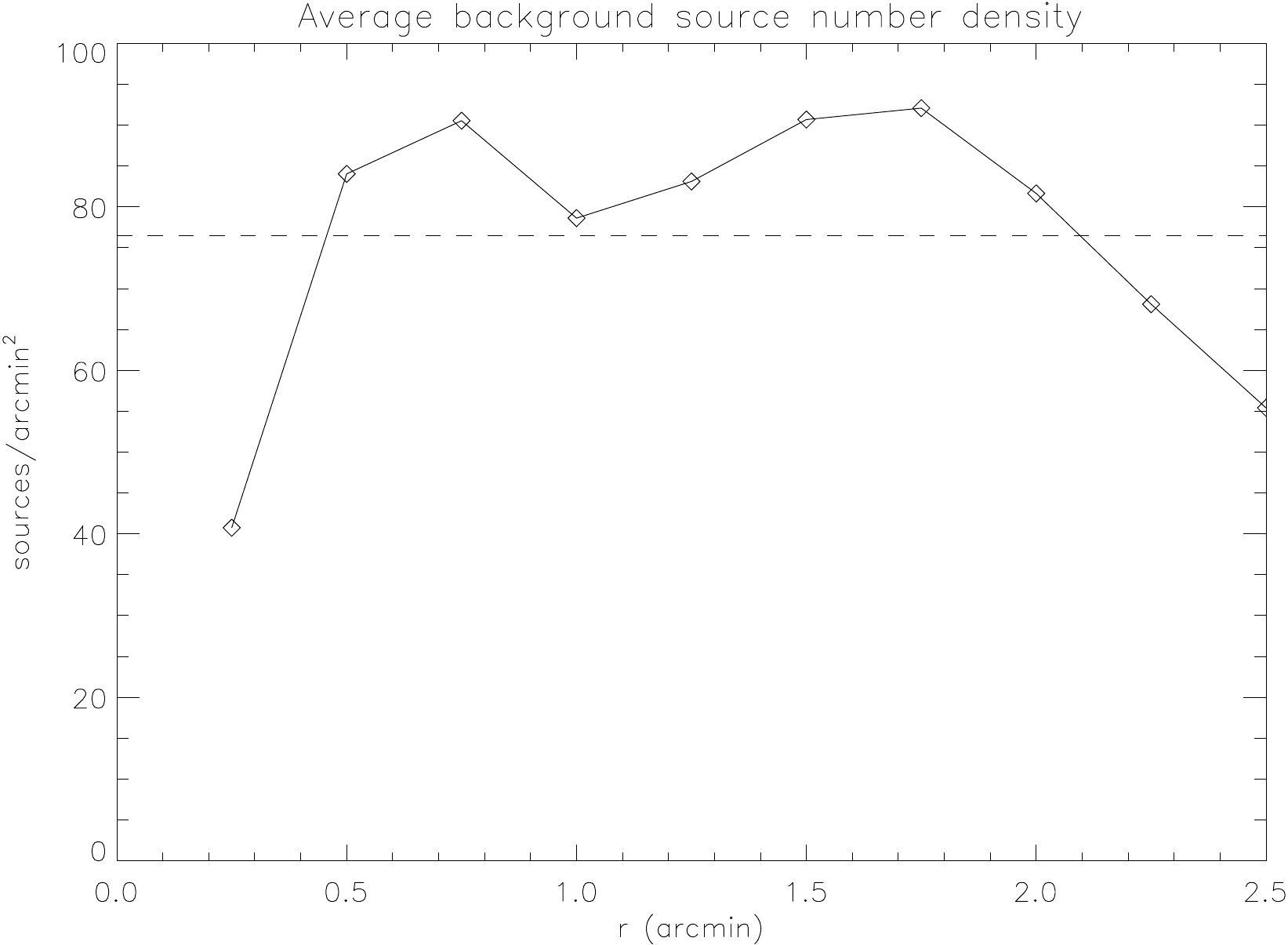}}
\caption{The figure shows the number density of background sources
  used in our flexion study (averaged over a circular region) plotted
  as a function of distance from the center of the image. The dotted
  line shows the number average number density over the entire
  image. There is an apparent downturn in the number density at large
  r. This downturn occurs simply because we reach the edge of the
  field.}
\label{fg:source_count}
\end{figure}

Clearly, the under density seen in the center of the convergence map
in Figure~\ref{fg:kappa_f} should be considered as resulting from
noise, rather than a real feature of the cluster, as the central
region appears to be entirely noise-dominated. This is most likely due
to the fact that there are fewer background sources found in this
region. Figure~\ref{fg:source_count} shows the average number density
of background sources plotted as a function of the radius over which
this number density is averaged. 

The shortage of sources in the central region results from our masking
scheme, and affects our flexion reconstruction in two ways. Firstly,
it means that a foreground galaxy in the center of the cluster is less
likely to have the required 3 nearby background sources, and thus
fewer of the central cluster members will be included in the
analysis. Secondly, those that are included will generally be fit
using fewer data points than the outlying cluster members, and thus
these fits will have larger associated errors.

Thus, the under density seen in the center of the flexion
reconstruction should not be believed. However, in the outlying
regions of the image, the noise is seen to drop significantly, and the
substructures seen in these regions appear to be real features.

\subsection{Galaxy-Galaxy Flexion Signal}

\begin{figure}[h]
\centerline{\includegraphics[angle=0,height=2.5in]{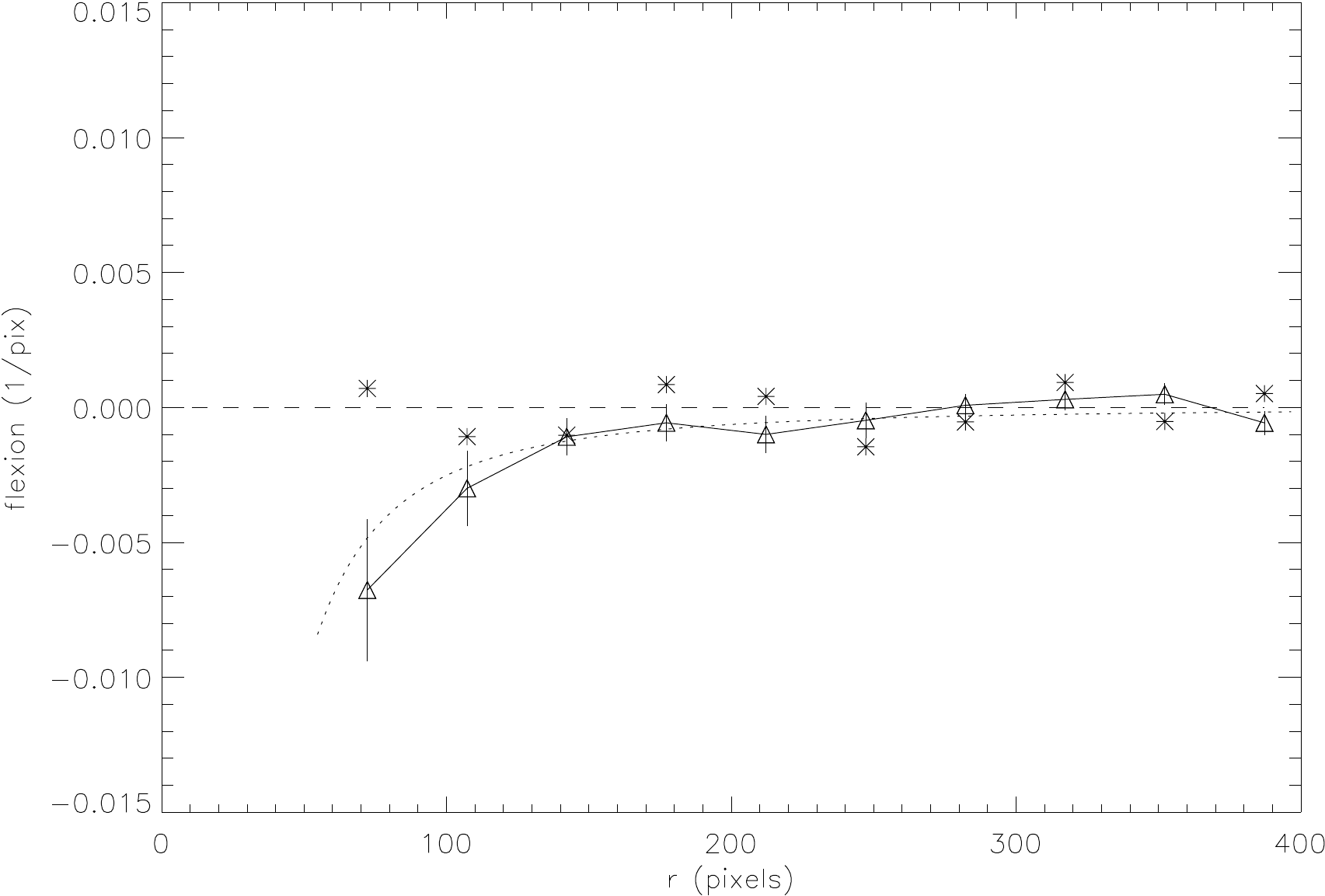}}
\caption{The mean galaxy-galaxy flexion signal of cluster
  galaxies. The B-mode signal is plotted as unconnected points, and
  is consistent with zero. The dotted line represents the expected
  signal for an isothermal sphere with a velocity dispersion
  $\sigma=295\ km/s$.}
\label{fg:gg}  
\end{figure}

In addition to a large-scale map of the cluster, we have also
generated a composite circular profile of the cluster member galaxies
via flexion measurements. Figure \ref{fg:gg} shows the galaxy-galaxy
(first-) flexion averaged over all background-foreground pairs meeting
our selection criteria.

We find that the mean cluster member can be fit well by an isothermal
sphere with $\sigma_v=295\pm 40\ km/s$.  Given the relatively large
uncertainty in velocity, and the relatively narrow scatter in the
magnitude of the cluster members ($\sigma_r\simeq 1$), we are unable
to effectively split the cluster members into subgroups.  However, we
can compare this result to that expected from the Faber-Jackon
relation.

The mean absolute magnitude of the sample is approximately -21.5 in
the R-band.  Taking the canonical Faber-Jackson relation:
\begin{equation}
\sigma_v=220\ km/s \left(\frac{L}{L_\ast}\right)^{0.25},
\end{equation}
we find an expected velocity dispersion of approximately 390 km/s, in
accord with our measurements.

L06 (Table 3) compute a best estimate RMS for several member galaxies
($\sim 200 km/s$) and the Brightest Cluster Galaxy ($\sim 500 km/s$).
Our mean flexion estimate falls squarely in the middle of this
distribution. 

We also note that the mean velocity dispersion found when computing
the parametric flexion reconstruction ($321 km/s$) is consistent with
that derived from the galaxy-galaxy lensing study, within the error
bars of the latter.

\section{Summary}

We have used a shapelet-based shear measurement technique to create a
non-parametric mass reconstruction of the galaxy cluster Abell
1689. Using only weak lensing data, we found significant ellipticity
and substructure in the cluster. Combining the weak lensing data with
strong lensing data from L06 improved the resolution of the mass map,
and increased the slope of the cuspy central density profile. The
combined analysis also identified a secondary dark matter clump found
by L06. Using an entirely new and independent flexion analysis, we
were able to verify the position of this clump, and other
substructure, via a parametric reconstruction of the cluster mass,
modeling each of the cluster members as a singular isothermal
sphere. The substructure observed in the flexion reconstruction is
well-correlated with the locations of groups of cluster galaxies.

We have also used flexion data to probe the halos of individual
cluster galaxies. Using a similar parametric reconstruction, we
measure a highly significant ($\sim 13\sigma$) galaxy-galaxy flexion
signal. In agreement with previous, non-flexion measurements, we find
a mean velocity dispersion of $295 \pm 40\ km/s$ for the cluster
galaxies.

\acknowledgements The authors would like to thank David Bacon,
Sanghamitra Deb, John Parejko, Lindsay King, and Marceau Limousin for
useful discussions. This work was supported by NASA ATP NNG05GF61G and
HST Archival grant 10658.01-A. AL is supported by a BP/PPARC Dorothy
Hodgkin Postgraduate Award.

\appendix

\section{Induced Flexion Due to the Point Spread Function}
\label{sec: derivation}

From equations 43-45 in GL07, we have the relations:

\begin{eqnarray}
  Q_{ijk}=Q_{ijk}^{(0)}+P_{ijk},\nonumber\\
  Q_{ijkl}=Q_{ijkl}^{(0)}+dP_{ijkl},
\end{eqnarray}

where 
\begin{equation}
  dP_{ijkl} \propto P_{ijkl} \propto a_{PSF}^4
\end{equation}

and 
\begin{equation}
  Q_{ijkl}^{(0)} \propto a_{source}^4.
\end{equation}

Now 
\begin{eqnarray}
  {\cal F} &\sim& \frac{Q_{ijk}}{Q_{ijkl}}\nonumber\\ &=&
  \frac{Q_{ijk}^{(0)}+P_{ijk}}{Q_{ijkl}^{(0)}+dP_{ijkl}}\nonumber\\
  &=& \frac{Q_{ijk}^{(0)}}{Q_{ijkl}^{(0)}}
  \frac{Q_{ijkl}^{(0)}}{Q_{ijkl}^{(0)}+P_{ijkl}} +
  \frac{P_{ijk}}{dP_{ijkl}} \frac{dP_{ijkl}}{Q_{ijkl}^{(0)}+dP_{ijkl}}
  \nonumber\\ &\sim& {\cal F}^{(0)}
  \frac{a_{source}^4}{a_{source}^4+a_{PSF}^4} + {\cal F}_{PSF}
  \frac{a_{PSF}^4}{a_{source}^4+a_{PSF}^4}.
\end{eqnarray}

The former term in the above expression will dominate for $a_{source}>
a_{PSF}$, and the coefficient of the ${\cal F}^{(0)}$ term will
approach unity as $a_{source}$ becomes large. In the case of small
$a_{source}$, the term in ${\cal F}_{PSF}$ will become important,
provided that ${\cal F}_{PSF}$ itself is significant.

\end{document}